# Quantitative Characterization of Brain Tissue Alterations in Brain Cancer Using Fractal, Multifractal, and IPR Metrics

Mousa Alrubayan, Santanu Maity, Prabhakar Pradhan

*Department of Physics and Astronomy*
*Mississippi State University, Mississippi State, MS, USA, 39762*

**Abstract:**

We studied the structural alterations between healthy and diseased brain tissues using a multiparametric framework combining fractal analysis, fractal functional transformation, multifractal analysis, and the Inverse Participation Ratio (IPR) analysis. Accurate characterization of brain tissue microstructure is crucial for early detection and diagnosis of cancer. By applying box-counting methods on brightfield microscopy images, we estimated the fractal dimension ($D_f$) and its logarithmic ($ln(D_f)$) and functional ($ln(D_{tf})$) forms to highlight spatial irregularities in the tissue architecture. While $D_f$ and $ln(D_f)$ exhibited long-tailed distributions distinguishing healthy from cancer tissues, $ln(D_{tf})$ provided significantly improved differentiation by emphasizing local structural variations. Additionally, multifractal analysis revealed broader $f(\alpha)$ vs $\alpha$ curves in cancerous samples, reflecting higher heterogeneity. IPR analysis based on light localization further demonstrated increased nanoscale variations in mass density, reflecting higher structural disorder in cancer tissues. Combining these complementary approaches creates a robust framework for measuring tissue complexity and holds great potential to improve microscopic diagnostic methods for brain cancer detection.

1. Introduction:

Brain cancer is widely recognized as one of the most challenging malignancies to diagnose and classify, due to the inherent complexity of its structure and composition [1]. Tumors of the brain often exhibit heterogeneous cellular organization, variable density, and irregular morphologies, all of which complicate their evaluation [1,2]. Conventional diagnostic approaches, including imaging modalities and histological inspection, play an essential role but are frequently limited in their ability to capture the subtle alterations that mark the earliest transitions from healthy to diseased tissue [3–5]. These limitations underscore the need for new quantitative strategies that can provide deeper insights into tissue organization and support more reliable diagnostic and prognostic assessments [6,7].

Fractal analysis has emerged as a powerful mathematical framework for studying structural irregularities in biological systems [8,9]. Unlike traditional linear measures, fractal geometry captures complexity across scales and provides a quantitative index of irregularity known as the fractal dimension [10]. This property makes it particularly suitable for analyzing cancer tissues, which often display disrupted architectures and non-uniform growth patterns [2,7]. Several studies have applied fractal dimension analysis to a range of malignancies, including breast, colon, and ovarian cancers [11–13], showing that cancerous tissues tend to exhibit higher fractal dimensions and broader variability than their normal counterparts [6,14]. Such findings demonstrate the promise of fractal measures as sensitive biomarkers of malignant progression [1,15].

Beyond single-value fractal descriptors, more advanced approaches have explored the statistical properties of fractal distributions. By considering the entire probability distribution of fractal indices, rather than only their mean, researchers have uncovered more nuanced markers of disease progression [16,17]. Transformations such as logarithmic scaling or functional transformations have been introduced to enhance contrast between different tissue states [15,16]. Additionally, statistical fitting methods, including polynomial approximations and Gaussian models, have been used to characterize distribution shapes, providing further tools for discriminating between healthy and diseased tissues [10,16].

Multifractal analysis extends this concept by evaluating the spectrum of fractal exponents within a tissue sample, thereby capturing variations in structural irregularity across different spatial scales [8]. Cancerous tissues typically produce broader and less symmetric multifractal spectra, reflecting their higher degree of heterogeneity and structural disorder [16,18]. This multifractal perspective provides a more comprehensive understanding of tumor complexity and offers complementary information to single-valued fractal measures [9].

Another promising quantitative tool is the inverse participation ratio (IPR), which originates from studies of disorder in weak disorder systems but has recently been adapted for biomedical imaging. IPR provides a measure of how localized or dispersed intensity patterns are within an image. In the context of tissue, high IPR values indicate clustered or irregular distributions, whereas low values correspond to more homogeneous, uniform patterns [19,20]. Cancerous tissues generally show elevated IPR values, signifying localized structural disorder and irregular cell clustering [3,4]. Because IPR and fractal dimension capture different aspects of tissue organization, their combined use has the potential to improve sensitivity and robustness in cancer detection [15,20].

Despite the growing interest in fractal, multifractal, and IPR-based analyses, their application to brain cancer remains relatively limited compared to other cancer types [21,22]. This gap motivates further research aimed at applying and refining these methods to brain tissue, with the goal of identifying distinctive structural markers that can aid in early diagnosis and more precise classification of disease stages.

## 2. Methods:

### 2.1. Tissue Sample Preparation and Image Acquisition:

This study used tissue images that were taken from brain specimens prepared from TMA (tissue microarray) samples from Biomax.us. This array contains 24 cores, each 1.5 mm in diameter and 5 microns thick, and contains both healthy and cancerous brain tissue. The specimen was imaged using an Olympus BX61 brightfield microscope. From the captured images, the 10 best images representing healthy tissue and the 10 best images representing diseased tissue were selected based on image quality and clarity of tissue structure.

### 2.2. Optical Principles of Brightfield Imaging:

In this analysis, we used brightfield images of brain samples taken using an Olympus BX61 brightfield optical transmission microscope (Tokyo, Japan). Biological tissues have optical and refractive properties; that is, it have a refractive index *n* that depends on the composition and mass density of the sample [6]. In our previous study, we showed that as the mass density increases within a tissue segment, the refractive index changes, and consequently, the intensity of the light passing through that segment changes [6,7,11,15]. The relationship between transmission intensity and mass density is described in the following equation:

$$I_t \propto (n_{tissue+glass} - n_{air}) \propto changes\ in\ tissue\ mass\ density. \qquad (1)$$

$I_t$ is the intensity of the light transmitted and recorded by the camera. $n_{tissue+glass}$ is the overall refractive index of the sample and glass, $n_{air}$ is the refractive index of air.

This behavior explains why microscopic images can be used to analyze the internal structure of a sample by examining its optical properties. We aimed to establish a connection between the variation in mass density and transmission intensity. Analyzing the transmitted intensity obtained from the tissue sample provides a clear way to estimate the fractal dimension, which is associated with mass density variation in tissue structure. Fractal dimension can be used as a biomarker in our further studies.

### 2.3. Box-Counting for Fractal Dimension $D_f$ Calculation in Brain Samples:

The goal of this section is to extract a quantitative measure of the structural complexity of brain tissue by converting the microscopic image into a fractal map that reflects the density distribution within the sample. The fractal dimension $D_f$ is used for this purpose [1,23], which represents a measure of the complexity or roughness of the structure at the fine-scale level. Optical transmission microscope images were first converted to grayscale, where each pixel is represented by a value between 0 (black) and 255 (white). Each image was then resized to a fixed size of 512 x 512 pixels to standardize processing across different samples. Finally, each image was divided into small squares (boxes) of 32 x 32 pixels. Since the total image size is 512 pixels, each image contains $(512/32)^2$ = 16 x 16 = 256 sub-boxes [6,7].

Within each sub-box, a box-counting method is applied. The square is covered with a grid of scale ε (2, 4, 8, 16, etc.). The number of boxes, $N(\varepsilon)$, containing pixels (i.e., non-empty areas), is calculated, and the previous step is repeated for multiple scales of ε. The relationship between $ln(N(\varepsilon))$ and $ln(1/\varepsilon)$ is plotted, and the slope of the line is extracted, which equals the value of the fractal dimension [1,6,10]. That is,

$$D_f = \ln(N(\varepsilon))/\ln(1/\varepsilon). \qquad (2)$$

In cancerous tissue, $D_f$ is often higher due to the random distribution and heterogeneous density of cellular components [6,1,15,11].

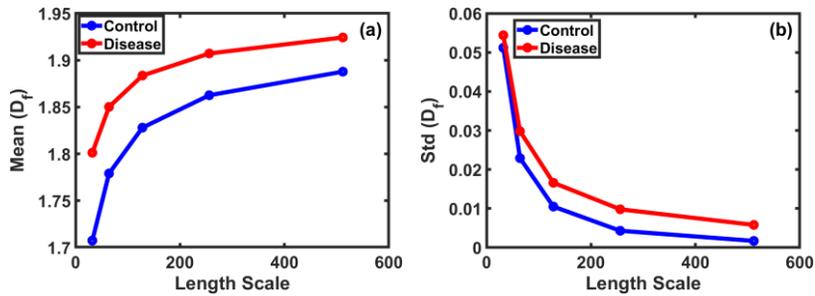

**Figure 1:** The relationship between the length scale used to calculate the fractal dimension ($D_f$) and the mean **(a)** and standard deviation **(b)** values for the control and disease samples. We observe that the mean gradually increases with increasing length scale in both groups, with the difference remaining clear across all scales. In contrast, the standard deviation decreases significantly with increasing length scale, indicating a loss of fine spatial detail when using large scales.

A complete map was generated for each image containing *256 $D_f$* values (one for each box). The mean $D_f$, the full probability distribution ($P(D_f)$) [6,7]. We studied the effect of varying length scale (32, 64, 128, 256, 512) on both the mean and standard deviation of the $D_f$ values in both control and disease samples. As shown in Figure 1(a), the mean $D_f$ gradually increases with increasing length scale for both groups, and the difference between control and disease remains significant at all scales [6]. However, figure 1(b) shows that the standard deviation decreases with increasing length scale, indicating a loss of fine spatial detail when using a very large scale. Based on these observations, the 32×32 length scale was chosen because it strikes an optimal balance [23] between preserving the contrast between the two groups (control and disease) while minimizing the loss of fine detail.

Similarly, the effect of the threshold ratio used to convert images to binary was analyzed on both the mean and standard deviation of $D_f$ values. As shown in Figures (2a) and (2b), we find that increasing the threshold leads to a gradual decrease in the mean $D_f$ with a significant increase in the standard deviation. It is noted that the difference in the mean $D_f$ between the control and disease samples reaches its maximum at a threshold ratio of 65% [15], while the largest difference in the standard deviation values between the two groups appears at a threshold ratio of 45%.

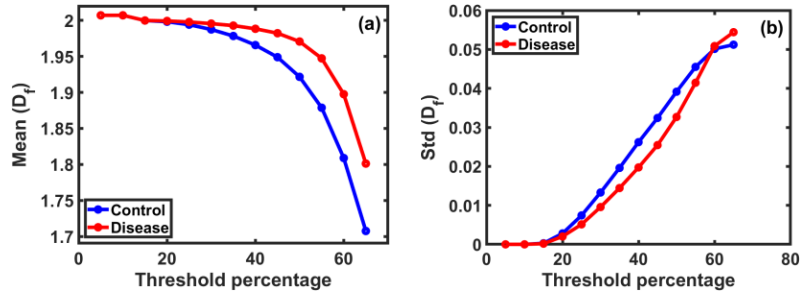

**Figure 2:** The effect of the binary threshold ratio on $D_f$ values for control and disease samples. In part **(a)**, we observe that the mean gradually decreases with increasing threshold ratio, with disease values consistently outperforming control values up to *65%*. In part **(b)**, the standard deviation shows a clear increase with increasing threshold, reflecting greater dispersion of values. It is worth noting that the maximum difference between the means appeared at the *65%* threshold, while the maximum difference in STD was at 45%. These ratios helped in choosing the optimal threshold for binary image analysis.

## 2.4. Quantifying Structural Heterogeneity with Multifractal Spectra:

In this part of the study, multifractal analysis was applied to tissue images to more accurately characterize the structural variation within brain tissue. This analysis is an effective tool for detecting subtle changes in tissue structure, as it can distinguish complex patterns that may not be apparent with single-fractal analysis [9,23].

Each image was divided into small squares, and the pixel intensities within each square were calculated. Using these intensities, the local probabilities for each square were calculated, which were used to calculate the f(α) spectrum using the Chhabra-Jensen method. This spectrum represents the distribution of the local exponents α and the associated fractal dimensions f(α), enabling analysis of the full distribution of local exponents [10,15,8].

The shape of the f(α) spectrum for each image was analyzed, focusing on the spectrum's width, symmetry, and peak location. Cancerous tissue is expected to exhibit a broader and less symmetrical multifractal spectrum compared to healthy tissue [8,9], indicating greater structural variation and irregularity in tissue structure.

To calculate the multifractal spectrum *f(α)*, a standard equation based on the probability distribution within squares was applied:

$$f(\alpha_Q) = Q \times \alpha_Q - \tau_Q = \sum_{i=1}^{n\varepsilon}(\mu_{i(Q,\varepsilon)} \times \ln(\mu_{i(Q,\varepsilon)}))/\ln \varepsilon \qquad (3)$$

where Q is the power index, tested in the range from -10 to +10, $\mu_i(Q,\varepsilon)$ represents the local probability of a pixel being located within the $i^{th}$ square at scale ε [8,23], derived from the tissue density distribution, $n_\varepsilon$ is the total number of boxes in scale ε, $\tau_Q$ is the scaling exponent associated with each value of Q, and $\alpha_Q$ is the local exponent associated with Q, calculated from the gradient of the $\tau_Q$ curve. This method shows high effectiveness in analyzing local differences in tissue structure, enabling the detection of subtle differences between healthy and cancerous tissue.

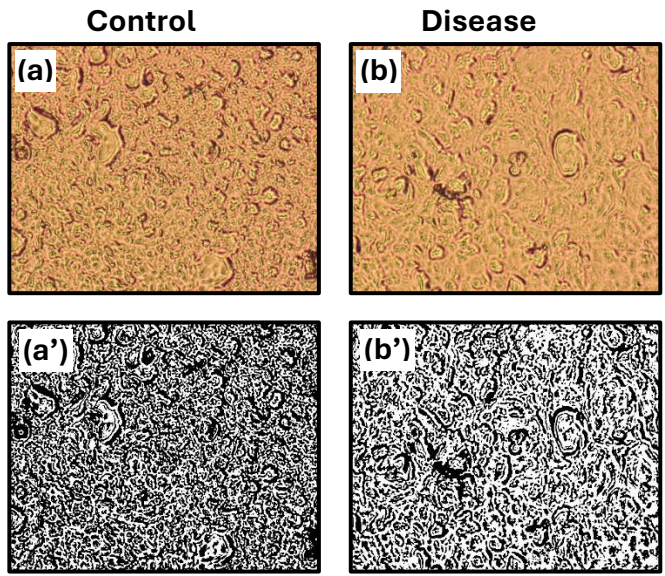

**Figure 3:** Brightfield microscopic images of brain tissue from **(a)** control and **(b)** disease samples. Below **(a', b')** show the same samples converted to binary images using a fixed threshold, allowing structural components to be extracted and analyzed fractionally via box-counting.

## 2.5. Functional Transformation of Fractal Dimension for Improved Tissue Differentiation:

After calculating the fractional dimensions $D_f$ using the box-counting method, it was observed that the $D_f$ distribution alone did not provide sufficient discrimination between healthy and diseased samples, especially in cases where the values were close together. To expand the scope of the analysis and improve the differentiation ability, a functional transformation was applied to $D_f$ [15].

$$D_{tf} = \frac{D_f}{D_{fmax} - D_f} \tag{4}$$

where $D_{fmax}$ is the maximum fractional dimension recorded across all images used [6,15]. This transformation contributes to changing the properties of the data distribution, increasing its variance and broadening the right tail of the distribution, making the differences between tissues more apparent when graphically represented [15]. This transformed variable is subsequently used to analyze its probability distribution and test how closely it approximates a Gaussian or lognormal distribution [15,23].

## 2.6. Statistical Characterization of Fractal Distributions:

The probability distributions of the variables $D_f$, $ln(D_f)$, $D_{tf}$, and $ln(D_{tf})$ were represented using histograms to study the overall shape of the distribution and evaluate its ability to distinguish between healthy and diseased tissue. To improve the appearance of the curves and reduce the influence of random fluctuations, a polynomial approximation was applied to the original curves. Gaussian approximation was then applied to the resulting distributions to extract their statistical parameters, such as the maximum value (peak) and standard deviation [15].

These parameters help assess the extent to which the distribution conforms to a Gaussian or log-normal shape, which is used as an indicator of the regularity of the spatial structure in the tissue [15]. It was observed that the Gaussian distribution tends to be sharply symmetric, while the log-normal distribution indicates higher variance and irregularity, which is consistent with the morphological features of cancerous tumors [10,15].

## 2.7. IPR-Based Analysis of Local Density and Structural Disorder in Tissue Images:

In this section, an inverse participation ratio (IPR) analysis was applied to brain tissue images to explore changes in nanoscale structural disorders in heterogeneous dielectric media, such as biological tissues, and to reveal the complexity of the tissue spatial structure. This analysis is based on calculating the degree of density localization or dispersion within the image and is a useful indicator for detecting random and irregular changes associated with cancerous growth [2,15,24].

IPR analysis compares IPR values between control and diseased tissues using the mean and standard deviation. Details about the IPR analysis are described elsewhere [2].

In a voxel of cells, changes in mass density $\rho(x,y)$ linearly depend on refractive index variation. Consequently, the transmission intensity is also linearly proportional to the mass density variation. The following equation represents the relationship between these parameters.

$$I(x,y) \propto \rho(x,y) \propto n(x,y) \qquad (5)$$

Here $I(x,y)$ denotes the transmission intensity measured from the sample, while $n(x,y)$ represents the refractive index variation.

Variation in the refractive index provides an optical lattice potential $\varepsilon_i$, which can be expressed as

$$\varepsilon_i = \frac{dn(x,y)}{n_0} \propto \frac{dI(x,y)}{I_0} \qquad (6)$$

Anderson's tight-binding model (TBM) showed that light can become localized in disordered cellular structures [21]. In this framework, biological tissues are viewed as optical lattices, with detailed descriptions provided elsewhere [15,24].

The tight-binding lattice model provides the Hamiltonian for a disordered system. For optical lattice, the Hamiltonian can be expressed as

$$H = \sum \varepsilon_i |i><i| + t \sum \langle ij \rangle \, (|i><j| + |j><i|) \qquad (7)$$

Here, $\varepsilon_i$ is the optical potential energy of eigenfunctions at the ith lattice site. $\langle ij \rangle$ representing the nearest-neighbor hopping interaction between these two lattice sites, and t is the overlap integral. $|i>$ and $|j>$ are the optical eigenfunctions at the $i^{th}$ and $j^{th}$ lattice sites.

In IPR analysis, mean IPR is computed using the following expression,

$$\langle IPR \rangle_{L \times L} = \frac{1}{N} \sum_{i=1}^{N} \int_0^L \int_0^L E_i^4(x,y) dx dy \qquad (8)$$

Mean IPR provides a quantitative measure of the degree of structural disorder. N defines the total number of optical eigenfunctions in the lattice of size L×L.

IPR value is linearly related to the refractive index fluctuations $<dn>$ and correlation length $l_c$, as described by the following equation

$$\langle IPR \rangle_{L \times L} \sim <dn>.l_c \qquad (9)$$

Additionally, IPR provides a quantitative measure of the structural disorder parameter $L_{d-ipr}$, which is also linearly dependent on it.

By combining the equations above, the following relationship can be derived.

$$\langle IPR \rangle_{L \times L} \propto L_{d-ipr} \sim <dn>.l_c \qquad (10)$$

$$STD(IPR)_{L \times L} \propto L_{d-ipr} \sim <dn>.l_c \qquad (11)$$

These quantifications provide a robust framework for estimating IPR parameters, enabling the detection of structural disorder during disease progression through the calculation of the mean and standard deviation of IPR.

To ensure that the chosen length scale is suitable for detecting spatial variations in tissue structure, we evaluated the effect of using different matrix sizes (2×2, 4×4, 8×8, 16×16, and 32×32) [15] on the resulting IPR values. As shown in the accompanying plots, smaller length scale such as 2×2 resulted in IPR values that were very close between control and disease groups, making it difficult to distinguish between the two. This is likely due to the extremely localized scale, which does not capture the broader structural patterns in mass density.

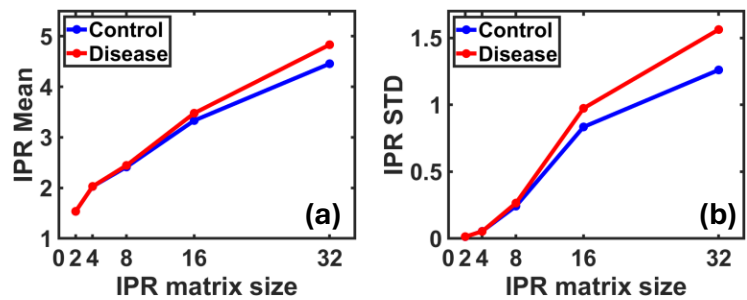

**Figure 4:** The relationship between the size of the matrix used to calculate the IPR coefficient and the mean **(a)** and STD **(b)** values for control and disease samples. We note that using small length scales such as 2×2 and 4×4 gives low and close values between the two groups, making it difficult to distinguish between them. As the length scale increases to 32×32, the differences become more pronounced, especially in disease samples, both in the mean and STD, enhancing the IPR's ability to distinguish between healthy and diseased microstructure.

However, as the length scale increased, the IPR values for the disease group began to diverge more clearly from the control group. At a length scale of 32×32, the difference became most prominent [4], both in terms of the mean, figure 4 (a), and the standard deviation, figure 4 (b). This larger scale allowed the analysis to capture more meaningful patterns in local density clustering and heterogeneity, which are characteristics commonly associated with cancer progression. Therefore, 32×32 was selected as the optimal length scale for the IPR computation in this study [15], as it provides better discriminatory power and highlights the potential of IPR as a promising marker for tissue abnormality.

## 3. Results:

### 3.1 Distribution of Fractal Metrics ($D_f$, $ln(D_f)$, $ln(D_{tf})$):

The probability distribution of the fractal dimension coefficient $D_f$ shows a clear contrast between control and disease tissue. This distribution was calculated using a fixed box size of *32* pixels, a binary threshold of *0.65*, and the number of distribution bins set to *70* for the histogram.

In Figure 5, we show the probability distribution of the fractal dimension coefficient $D_f$ and its logarithmic transformation $ln(D_f)$ using the polynomial approximation to compare images of control and disease tissue. In Figure 5(a), the curve of the control group appears narrow and symmetric around a specific value, while the disease group curve is wider and slightly shifted to the right. This indicates greater variation in $D_f$ values, reflecting structural differences caused by the disease. In Figure 5(b), after applying the logarithmic transformation ($ln(D_f)$), the contrast between the two groups becomes more pronounced. The control group remains centered with a tight distribution, while the disease group curve becomes broader and less symmetric. This supports that $ln(D_f)$ enhances the visibility of changes in the tissue structure due to the disease. These differences suggest that both $D_f$ and $ln(D_f)$, when approximated using polynomial fitting, offer useful indicators for capturing pathological changes in tissue morphology.

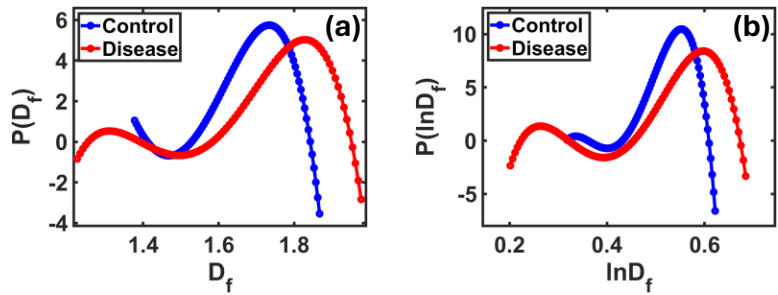

**Figure 5**: Polynomial approximation of $D_f$ and $ln(D_f)$ distributions for control and disease tissue. In **(a)**, the disease curve is broader and shifted right, indicating higher structural complexity. In **(b)**, the $ln(D_f)$ curve is wider and less symmetric, reflecting deeper morphological changes.

In Figure 6, we show the distribution of the $ln(D_{tf})$ coefficient using two different methods: the Polynomial approximation 6(a) and the Gaussian approximation 6(b). This coefficient, resulting from a nonlinear transformation of $D_f$, provides a stronger contrast between control and disease samples compared to the previous metrics. In Figure 6(a), we note that the curve for control samples appears almost symmetrical and centered around a clear value, while the curve for disease samples is broader and contains irregular undulations, indicating greater structural variations within the affected tissue. This variation in curve shape demonstrates the sensitivity of $ln(D_{tf})$ to detecting the diversity of microscopic details in tissues.

In Figure 6(b), using the Gaussian approximation, the distinction becomes even more obvious. The control group shows a sharper, more concentrated curve, while the disease group's curve is wider and shifted rightward, pointing to a broader spread in $ln(D_{tf})$ values due to disease-related texture changes. These observations confirm that $ln(D_{tf})$ is particularly effective in detecting fine compositional changes that may not appear with standard methods.

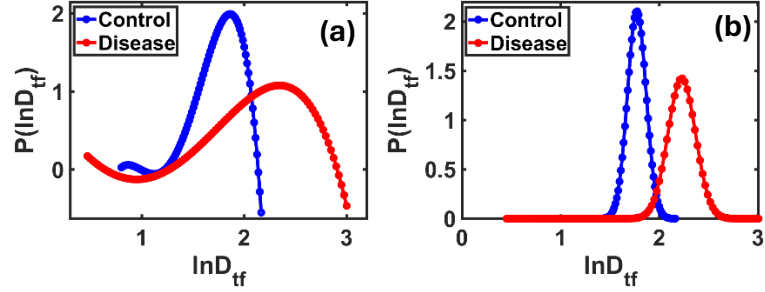

**Figure 6:** Polynomial **(a)** and Gaussian **(b)** approximations of the $ln(D_{tf})$ distribution for control and disease samples. The disease curve in (a) appears wider with irregular undulations, indicating higher structural complexity. In **(b)**, the disease curve is broader and right-shifted, confirming greater dispersion and morphological differences. The chi-square test for Gaussian fitting gives scores > 90%.

To complement these visual observations, Table 1 provides a numerical comparison of the three coefficients ($D_f$, $ln(D_f)$ and $ln(D_{tf})$) between the control and disease groups. The $D_f$ metric shows a moderate difference of 5.48% in the mean and 6.26% in the standard deviation, indicating some level of contrast using conventional analysis. The $ln(D_f)$ coefficient improves the separation with a 9.98% difference in mean and a slight change in standard deviation.

However, the $ln(D_{tf})$ coefficient clearly stands out, with a 25.69% difference in the mean and 47.82% in the standard deviation, highlighting its strength in detecting subtle histological changes between healthy and diseased brain tissue.

**Table 1:** Comparison of $D_f$, $ln(D_f)$ and $ln(D_{tf})$ between control and diseased samples in terms of mean and STD. $ln(D_{tf})$ stands out as the most discriminating indicator due to the large differences between the two groups.

| Fractal Metric | Mean | | | | STD | | | |
|---|---|---|---|---|---|---|---|---|
| | Control | Disease | Difference | Percentage % | Control | Disease | Difference | Percentage % |
| $D_f$ | 1.7075 | 1.8012 | 0.0936 | 5.48% | 0.0512 | 0.0544 | 0.0032 | 6.26% |
| $ln(D_f)$ | 0.5346 | 0.5879 | 0.0534 | 9.98% | 0.0308 | 0.0314 | 0.0007 | 2.15% |
| $ln(D_{tf})$ | 1.7728 | 2.2283 | 0.4555 | 25.69% | 0.1895 | 0.2802 | 0.0906 | 47.82% |

## 3.2 Multifractal Spectrum f(α) of Brain Tissue:

*f(α)* analysis was applied to the images generated from control and diseased samples using the Chhabra-Jensen method, and the results are shown in Figure 7. The curve illustrates the relationship between *f(α)* and α for both groups, revealing a clear difference in the overall shape of the distribution.

The control curve (blue) was more centered, with a smaller width and a balanced slope on both sides, reflecting a certain degree of structural organization in healthy tissue. The disease curve (red) was wider and less symmetrical, and extended toward higher α values, indicating a more complex and uneven distribution within the diseased tissue. The difference in the shape of the two curves reflects that brain tissue in the diseased state becomes more irregular and more heterogeneous in microstructure, especially as the disease progresses.

This result is consistent with what we observed for the $D_f$ and $ln(D_{tf})$ coefficients, but here it reflects changes more sensitively because it represents the entire distribution of local α indices within the sample, rather than just a single average value. Therefore, the analysis of *f(α)* provides an important additional dimension that helps in detecting subtle structural changes between samples.

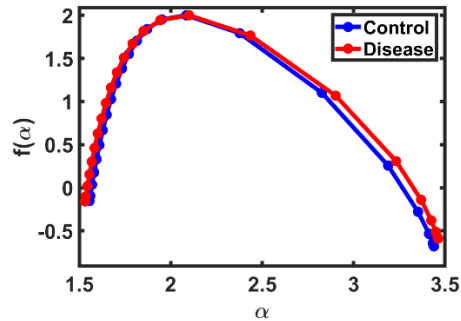

**Figure 7:** *f(α)* spectrum for control and disease samples. The curve for both groups shows multifractal behavior, with clear overlap and slight variation in spectrum width, indicating different structural complexity between tissues.

## 3.3. Identification of complex microscopic structures using IPR:

The Inverse Participation Ratio Index (IPR) was used to analyze the microstructural properties of brain tissue in the data images. This index allows assessing the level of localization of local variations within the image, reflecting the degree of "centering" or "scattering" in the structural distribution.

IPR values were extracted from the red channel using small-scale segmentation (length scale = 32), where eigenvector values were calculated and analyzed for each segment.

The results showed clear differences between control and diseased samples, indicating significant structural changes in the pixel density distribution in diseased tissue compared to healthy tissue, indicating increased structural disorder in cancer cells.

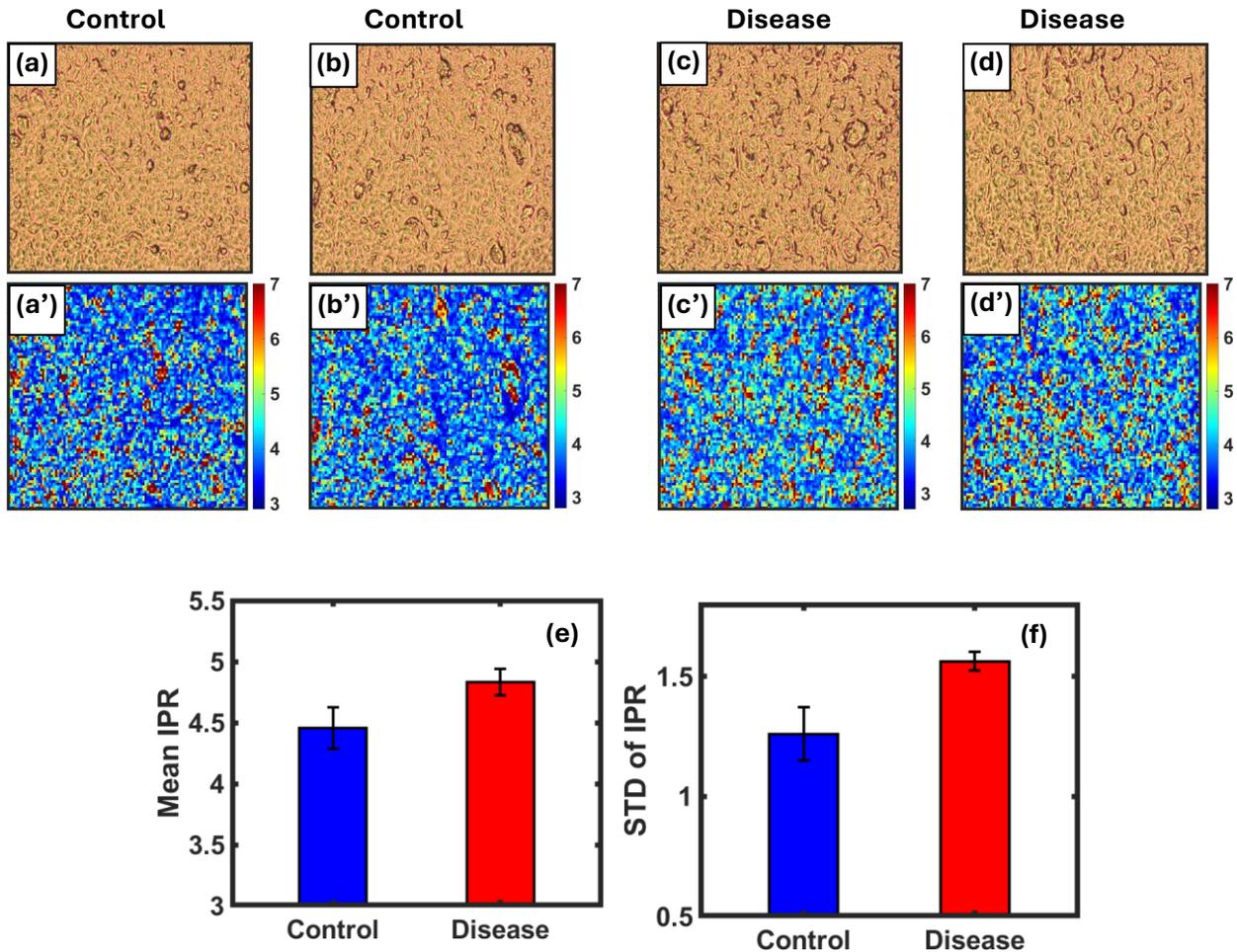

**Figure 8:** Brightfield images of brain tissue samples and their corresponding IPR heatmaps. Panels **(a)** and **(b)** represent control samples, while **(c)** and **(d)** show cancerous tissue. The lower row **(a′–d′)** illustrates the spatial distribution of the IPR values computed from the red channel using a 32×32 block size. Control samples exhibit more uniform and lower IPR values (shown in blue), indicating a more homogeneous structure. In contrast, the disease samples show scattered regions of high IPR (red/yellow), reflecting localized density concentration and greater structural irregularity.

**(e)** Mean and **(f)** standard deviation of the IPR coefficient for control and disease samples, with error bars representing standard deviation. The disease samples exhibit higher mean and standard deviation values, indicating greater structural dispersion in the tissue.

The standard deviation of the IPR values was 1.261 for healthy samples compared to 1.563 for diseased samples, a relative increase of approximately 24% (Figure 8(f)). The error bars highlight this variability, confirming greater diversity in the microstructure of diseased tissue.

Based on all the previous analyses, each parameter used in the study provided a different picture of the structural differences between healthy and diseased tissues, but some were more sensitive than others. The $D_f$ parameter showed a slight difference in distribution and standard deviation, but it was not sufficient on its own to clearly distinguish between the two groups. When we used $ln(D_f)$, the ability to separate the control from the disease improved, but the differences were still limited.

In contrast, $ln(D_{tf})$ had a very clear effect, whether in terms of the distribution expansion, the peak shift, or even the significant increase in the standard deviation, and it was one of the clearest indicators that differentiated the two groups. As for the $f(\alpha)$ analysis, it had strong added value because it did not limit itself to a single value but rather presented a complete distribution that illustrates the extent of structural diversity within each sample. The broader curve in the disease group gives a clear impression of the presence of different degrees of complexity within the same picture, which reflects the changes that occur in the cells as the disease progresses.

Finally, IPR analysis supported all our findings, providing direct numerical evidence of the scatter of images, showing that diseased samples are more concentrated in certain regions within the tissue, an indication of the internal structural chaos evident in conventional methods. In general, we can say that $ln(D_{tf})$, $f(\alpha)$, and IPR are the most powerful in terms of discrimination, and each of them gives a different value, but they ultimately reach the same conclusion: cancerous tissue has significant internal variation, greater structural complexity, and distributional properties that differ significantly from healthy tissue.

4. Discussion

This study analyzed structural differences between healthy and diseased brain tissues using advanced fractal and statistical indicators: $D_f$, $ln(D_f)$, $ln(D_{tf})$, $f(\alpha)$, and IPR. Each metric captured different aspects of tissue microstructure, and together they provided a comprehensive view of the spatial complexity linked to cancer progression [22,25].

The $D_f$ metric revealed noticeable, but insufficient, differences due to overlapping values, limiting its diagnostic utility [6,26].

Applying the $ln(D_f)$ transformation enhanced group separation by shifting central values and increasing statistical contrast [8,27], though improvements remained moderate.

In contrast, $ln(D_{tf})$ showed the most distinct separation. By compressing high values and expanding low ones, it captured fine-scale heterogeneity, with diseased tissues exhibiting a wider distribution and higher standard deviation.

The $f(\alpha)$ spectrum added further insight, showing wider curves in cancer samples — indicating structural complexity and heterogeneity not detectable by global averages. Although asymmetry was limited, the curve's breadth highlighted irregular cellular compositions in diseased tissues [10,17].

Finally, IPR effectively quantified local intensity clustering. Higher IPR values and standard deviations in cancer samples pointed to structural disorder and localized pixel intensity, emphasizing the heterogeneity within tumor tissue. This supports IPR as a complementary tool to fractal metrics in detecting early pathological changes [3,20,28].

Together, $ln(D_{tf})$, $f(\alpha)$, and IPR form a powerful combination for quantifying and detecting subtle structural differences between healthy and diseased tissue.

## 5. Conclusions

In this study, a set of structural and mathematical metrics were used to analyze microstructural differences between healthy and cancerous brain tissue. The results showed that some metrics, such as $ln(D_{tf})$, $f(\alpha)$ spectrum, and the IPR indicator, were more sensitive and efficient in detecting microstructural changes associated with the presence of a tumor than conventional metrics. These metrics provided an accurate quantitative description of the internal complexity and density distribution within the tissue, enabling clear discrimination between healthy and diseased samples even in the early stages of pathological change.

These results highlight the importance of combining fractal methods with spatial distribution metrics such as IPR to gain a deeper understanding of tissue structure. Each of the metrics studied provides a different perspective on the internal organization of the tissue, ranging from the density distribution ($D_f$ and $ln(D_f)$) to the complexity of local variation ($f(\alpha)$ and IPR). This multi-indicator approach not only reinforces the detection of cancer-related microstructural changes but also lays the groundwork for future diagnostic tools that rely on detailed tissue characterization through microscopic imaging.

**Acknowledgements:** This work was partially supported by the National Institutes of Health (NIH) under grant number R21 CA260147 to PP.